\documentclass[%
 aip,
 jmp,%
 amsmath,amssymb,%
reprint,%
]{revtex4-1}

\usepackage{graphicx}
\usepackage{dcolumn}
\usepackage{bm}
\usepackage[mathlines]{lineno}

\begin{document}

\title{Cluster Altered Magnetic and Transport Properties \\ in Eu Co-Doped Ge$_{1\textrm{-}x}$Mn$_{x}$Te}

\author{L.~Kilanski}
\email[Electronic mail: ]{kilan@ifpan.edu.pl}
\author{M.~G\'{o}rska}
\author{R.~Szymczak}
\author{W.~Dobrowolski}
\author{A.~Podg\'{o}rni}
\author{A.~Avdonin}
\author{V.~Domukhovski}
\affiliation{Institute of Physics, Polish Academy of Sciences, al. Lotnikow 32/46, 02-668 Warsaw, Poland}

\author{V.~E.~Slynko}
\author{E.~I.~Slynko}
\affiliation{Institute of Materials Science Problems, Ukrainian Academy of Sciences, 5 Wilde Street, 274001 Chernovtsy, Ukraine}

\date{\today}

\begin{abstract}

Magnetic and transport properties of Ge$_{1\textrm{-}x\textrm{-}y}$Mn$_{x}$Eu$_{y}$Te crystals with chemical compositions 0.041$\,$$\leq$$\,$$x$$\,$$\leq$$\,$0.092 and 0.010$\,$$\leq$$\,$$y$$\,$$\leq$$\,$0.043 are studied. Ferromagnetic order is observed at 150$\,$$<$$\,$$T$$\,$$<$$\,$160$\;$K. Aggregation of magnetic ions into clusters is found to be the source of almost constant, composition independent Curie temperatures in our samples. Magnetotransport studies show the presence of both negative (at $T$$\,$$<$$\,$25$\;$K) and linear positive (for 25$\,$$<$$\,$$T$$\,$$<$$\,$200$\;$K) magnetoresistance effects (with amplitudes not exceeding 2\%) in the studied alloy. Negative magnetoresistance detected at $T$$\,$$<$$\,$25 K is found to be due to a tunneling of spin-polarized electrons between ferromagnetic clusters. A linear positive magnetoresistance is identified to be geometrical effect related with the presence of ferromagnetic clusters inside semiconductor matrix. The product of the polarization constant and the inter-grain exchange constant, $JP$, varies between about 0.13$\;$meV and 0.99$\;$meV. Strong anomalous Hall effect (AHE) is observed for $T$$\,$$\leq$$\,$$T_{C}$ with coefficients $R_{S}$ independent of temperature. The scaling analysis of the AHE leads to a conclusion that this effect is due to a skew scattering mechanism.

\end{abstract}





\maketitle

\section{Introduction}\label{Introduction}

A novel class of spintronic devices based on semimagnetic semiconductors (SMCS's) attracted large interest in the past years.\cite{Ohno1998a} Many SMCS's ternary ferromagnetic compounds belonging to III-V and II-VI groups are nowadays intensively studied, because they combine magnetic properties useful for applications and controllable electronic, transport, and optical properties. Beyond them exists another interesting group of materials belonging to IV-VI SMCS's, such as (Ge,Mn)Te alloy, in which carrier controlled ferromagnetism with the Curie temperature, $T_{C}$, as high as 200$\;$K for $x$$\,$$\approx$$\,$0.08 was also observed.\cite{Fukuma2008a, Hassan2010a, Dietl2010a} High solubility of Mn (0$\,$$\leq$$\,$$x$$\,$$\leq$$\,$1) and high native hole concentrations ($n$$\,$$\approx$$\,$10$^{20}$$\,$$\div$$\,$10$^{21}$$\;$cm$^{-3}$) makes this material excellent for observation and controlling high temperature carrier induced ferromagnetism.\cite{Rodot1966a, Cochrane1974a} The magnetic properties of many different representatives of IV-VI SMCS's show a broad ability to control the type and the temperature of the magnetic transition.\cite{Kilanski2008a, Kilanski2009a, Kilanski2010b, Kilanski2012a}  \\ \indent Due to fundamental difficulties with obtaining room temperature ferromagnetism in SMCS's a new class of hybrid semiconductor - ferromagnetic metal systems is recently intensively developed and studied; see e.g. Refs$\;$\onlinecite{Wellmann1997a, Yuldashev2001a, Johnson2010a}. Development of GeTe based ferromagnetic composite materials\cite{Podgorni2012a, Kilanski2013b} creates possibilities to improve magnetic properties of the material and introduce new and interesting magnetotransport effects resulting from the presence of metallic grains inside the semiconductor matrix. \\ \indent In this paper, we present a detailed study of magnetic and transport properties of Ge$_{1\textrm{-}x\textrm{-}y}$Mn$_{x}$Eu$_{y}$Te mixed crystals. The main goal of this work is to study the possibilities of tuning the properties of the GeTe based composite by means of addition of two kinds of paramagnetic ions, i.e., manganese and europium. The magnetic properties of Ge$_{1\textrm{-}x\textrm{-}y}$Mn$_{x}$Eu$_{y}$Te crystals were studied, reporting ferromagnetism with $T_{C}$$\,$$\approx$$\,$120$\;$K (for $x$$\,$$=$$\,$0.073, $y$$\,$$=$$\,$0.03) and a noticeable contribution from antiferromagnetic superexchange interaction. That contribution reduced an effective magnetic moment of the material.\cite{Dobrowolski2006a, Dobrowolski2007a, Brodowska2008a} By addition of a few percents of Mn and Eu ions to the alloy we expect to enhance the strength of the long range magnetic interaction in the studied system and possibly increase its Curie temperature. We would also like to investigate the influence of the paramagnetic ion content on the magnetotransport of the alloy.

\section{Sample characterization}\label{SampleCharacterization}

Our Ge$_{1\textrm{-}x\textrm{-}y}$Mn$_{x}$Eu$_{y}$Te crystals were grown using the modified Bridgman method. The inclined crystallization front technique (applied for the first time by Aust and Chalmers to enhance the quality of aluminum \cite{Aust1958a}) was employed for the growth of these crystals. With this technique it is possible to decrease the number of crystal blocks in the grown ingot from a few down to a single one. The synthesis was done with the use of MnTe instead of metallic Mn. That eliminates the possibility of appearance of the metallic Mn precipitates in the as-grown crystals. Preliminary synthesis of EuTe prior to alloy growth is not necessary (due to its sufficiently low melting point equal to 826$^{\circ}$C). Therefore, metallic Eu was used. The crystal synthesis was performed at high temperatures equal to 1200-1250$^{\circ}$C. The applied growth temperature ensured an effective chemical bonding of tellurium (which acts like oxygen in oxidation) with manganese, preventing the appearance of metallic Mn precipitates. The growth material was prepared in a way ensuring that all cation elements are balanced by equal molar amount of tellurium. Each as-grown ingot was cut into slices about 1$\;$mm thick, perpendicular to the growth direction, prior to the further characterization. \\ \indent The chemical composition of the crystals is determined using the energy dispersive x-ray fluorescence (EDXRF) technique. The EDXRF analysis shows that the chemical composition of our samples changes continuously along the growth direction. Relative chemical content gradient within an individual sample does not change much (maximum relative change does not exceed 5\%). Two sets of samples in which only one type of substitutional elements changes (within the EDXRF experimental error) are selected for further studies. The first set of samples has Eu content changing in the range of 0.018$\,$$\leq$$\,$$y$$\,$$\leq$$\,$0.043 (while Mn content is almost constant, i.e., $x$$\,$$\approx$$\,$0.076$\pm$0.002) and the second one has Mn content varying between 0.041$\,$$\leq$$\,$$x$$\,$$\leq$$\,$0.092 (while Eu content is kept constant i.e. $y$$\,$$\approx$$\,$0.015$\pm$0.005). \\ \indent The crystallographic quality of the Ge$_{1\textrm{-}x\textrm{-}y}$Mn$_{x}$Eu$_{y}$Te crystals is studied using the x-ray diffraction (XRD) technique. All the XRD measurements are done at room temperature using the Siemens D5000 diffractometer. The obtained results show that all the studied samples are single phased and crystallize in the NaCl structure with rhombohedral distortion in the (111) direction. It should be emphasized, that within the experimental accuracy, no other phases are observed. The diffraction patterns are fitted with the use of Rietveld method in order to calculate the crystallographic parameters of our samples. The resulted lattice parameters are similar to their nonmagnetic equivalent, i.e., GeTe with $a$$\,$$=$$\,$5.98$\;$$\textrm{\AA}$ and $\alpha$$\,$$=$$\,$88.3$^{\circ}$.\cite{Galazka1999a} However, since changes in the lattice parameters were very small (maximum relative change of about 1\%), it was not possible to determine any obvious chemical trends in their values.

\section{Magnetic properties}\label{MagneticProperties}

\subsection{Low field results}

\begin{figure}[t]
  \begin{center}
    \includegraphics[width = 0.50\textwidth, bb = 10 40 582 386]
    {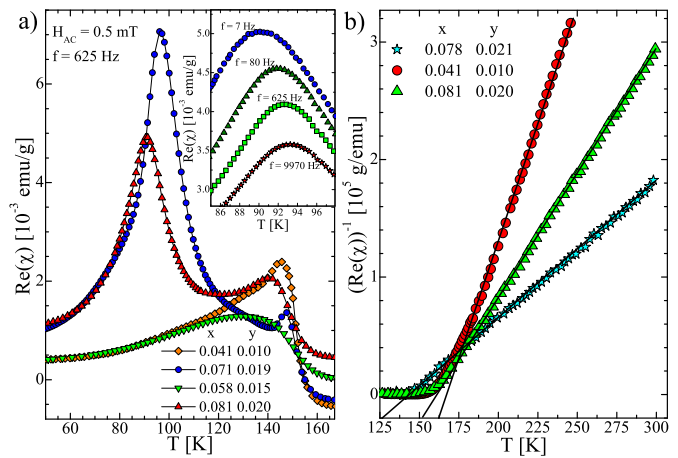}\\
  \end{center}
  \caption{\small Results of the ac susceptibility measurements including: a) low temperature region of real Re($\chi$) component of the ac linear magnetic susceptibility and the frequency shifting of Re($\chi$) (inset) and b) the inverse of the magnetic susceptibility for the selected Ge$_{1\textrm{-}x\textrm{-}y}$Mn$_{x}$Eu$_{y}$Te samples with different chemical compositions (shown in legend).}
  \label{FigACSuscVsTemp}
\end{figure}

\begin{figure}[t]
  \begin{center}
    \includegraphics[width = 0.42\textwidth, bb = 0 40 650 580]
    {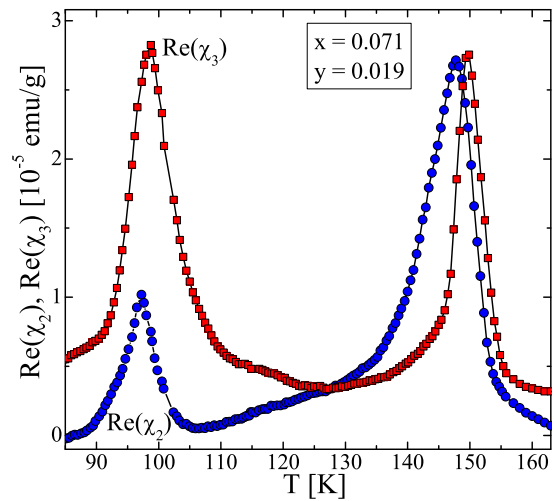}\\
  \end{center}
  \caption{\small The temperature dependence of the second and third harmonic ac susceptibility for a selected Ge$_{0.910}$Mn$_{0.071}$Eu$_{0.019}$Te crystal.}
  \label{FigACHarmVsTemp}
\end{figure}

\begin{figure}[t]
  \begin{center}
    \includegraphics[width = 0.42\textwidth, bb = 0 40 650 580]
    {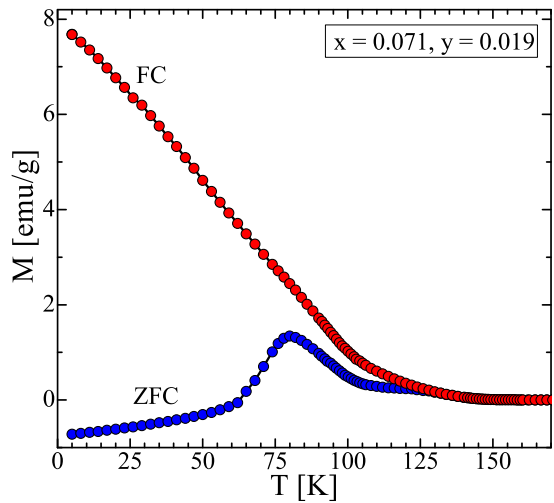}\\
  \end{center}
  \caption{\small The temperature dependence of the magnetization for a selected Ge$_{0.910}$Mn$_{0.071}$Eu$_{0.019}$Te crystal.}
  \label{FigFCZFCMagnVsTemp}
\end{figure}

The magnetic properties of Ge$_{1\textrm{-}x\textrm{-}y}$Mn$_{x}$Eu$_{y}$Te crystals are studied by means of static and dynamic magnetometry using the LakeShore 7229 susceptometer/magnetometer and Quantum Design Magnetic Property Measurement System (MPMS) XL-7 magnetometer systems. \\ \indent Mutual inductance method employed into the LakeShore 7229 ac susceptometer/magnetometer system is used in order to determine the temperature dependencies of both real Re($\chi$) and imaginary Im($\chi$) parts of the linear part of the ac susceptibility. The imaginary part of the ac susceptibility is close to zero and temperature independent for all our samples. Typical results in the form of the temperature dependence of the ac susceptibility for a few selected samples with different chemical compositions are presented in Fig.$\;$\ref{FigACSuscVsTemp}a.
An increase of Re($\chi$) at temperatures around 140$\,$$\div$$\,$150$\;$K is observed in all our samples. An appearance of a second peak in Re($\chi$) at temperatures around 80$\,$$\div$$\,$90$\;$K is also observed in most of Ge$_{1\textrm{-}x\textrm{-}y}$Mn$_{x}$Eu$_{y}$Te crystals. It indicates that two magnetic transitions are observed in our system. However, more detailed measurements need to be performed in order to determine the nature of the observed magnetic phases. \\ \indent The real part of the linear ac susceptibility is measured also in the temperature range significantly above $T$$\,$$=$$\,$150$\;$K, i.e., in the paramagnetic region (see Fig.$\;$\ref{FigACSuscVsTemp}b). The (Re($\chi$))$^{-1}$(T) dependencies in the case of all the Ge$_{1\textrm{-}x\textrm{-}y}$Mn$_{x}$Eu$_{y}$Te crystals are nearly linear up to maximum studied temperature, i.e., around 320$\;$K. It is a direct signature of the absence of magnetic clusters showing magnetic order at temperatures higher than 150$\;$K, in particular the MnTe antiferromagnetic clusters with the Neel temperature equal to 720 K.\cite{Cochrane1974a}  \\ \indent Measurements of temperature dependence of Re($\chi$) at four different frequencies of an alternating magnetic field $f$$\,$$=$$\,$7, 80, 625, and 9980$\;$Hz are also done. No frequency shifting of the susceptibility peak at $T$$\,$$\approx$$\,$140$\div$150$\;$K is observed in the studied samples. Only the maxima of the Re($\chi$)($T$) dependence near $T$$\,$$\approx$$\,$80$\div$90$\;$K shift on the temperature scale with an increasing $f$ (see an exemplary result in the inset to Fig.$\;$\ref{FigACSuscVsTemp}a). The frequency dependent data in the vicinity of maxima at $T$$\,$$\approx$$\,$90$\;$K are analyzed using the $R_{M}$ parameter proposed by Mydosh\cite{Mydosh1994a} and expressed in the following form:
\begin{equation}\label{EqMydosh}
    R_{M}=\frac{\Delta T_{F}}{(T_{F})\Delta log{(f)}},
\end{equation}
where $T_{F}$ represents the freezing temperature at a frequency $f$. The $T_{F}$($f$) values are determined as the position of the maximum in the Re($\chi$)($T$) dependence measured at the frequency $f$. The calculated values of the $R_{M}$ factor change between 0.005 and 0.018 without any signature of being chemical composition dependent. Our $R_{M}$ values are close to those reported in the case of spin-glass systems, e.g. CuMn with $R_{M}$$\,$$\approx$$\,$0.005 or AuFe with $R_{M}$$\,$$\approx$$\,$0.010.\cite{Mydosh1994a} The obtained results indicate that our Ge$_{1\textrm{-}x\textrm{-}y}$Mn$_{x}$Eu$_{y}$Te samples show paramagnet-ferromagnet transition at temperatures about 150$\;$K. Moreover, the second magnetic transition detected at $T$$\,$$\approx$$\,$80$\div$90$\;$K can be identified as an reentrant spin-glass freezing of magnetic moments. The observed magnetic transitions are related to the clusters present in our samples. Therefore, the magnetic state present at temperatures lower than 85$\;$K will be referred to the reentrant cluster-glass state and the magnetic order observed from 85$\;$K up to 140$\;$K will be referred to cluster-ferromagnetism. \\ \indent We also performed measurements of temperature dependence of higher harmonic susceptibilities in all Ge$_{1\textrm{-}x\textrm{-}y}$Mn$_{x}$Eu$_{y}$Te crystals. The obtained results (see Fig.$\;$\ref{FigACHarmVsTemp}) show that in the vicinity of both magnetic transitions, i.e., at $T$$\,$$\approx$$\,$80$\div$90$\;$K and $T$$\,$$\approx$$\,$140$\div$150$\;$K we observe the appearance of peaks in temperature dependencies of both second and third harmonic susceptibility curves.
It should be noted, that Re($\chi_{2}$)($T$) curve shows divergent behavior even at temperature of about 85$\;$K, at which a transition to a reentrant cluster-glass state is observed. It is a clear signature, in agreement with previous results, that ferromagnetic interactions are strong in this system. We can also clearly see, that in the vicinity of a cluster-glass transition the peak in Re($\chi_{3}$)(T) curve is higher than the one in Re($\chi_{2}$)(T) dependence. It indicates that the magnetic frustration observed near the peak at $T$$\,$$\approx$$\,$80$\div$90$\;$K attributed to a cluster-glass transition is stronger than the frustration near the peak at 140$\div$150$\;$K, supporting our earlier interpretation, that we do observe a cluster ferromagnet - reentrant cluster-glass transition. \\ \indent The temperature dependencies of both zero-field-cooled (ZFC) and field-cooled (FC) magnetization $M$ are measured using a SQUID magnetometer at the temperature range between 4.5 and 300$\;$K using constant magnetic field $B$$\,$$=$$\,$0.5$\;$mT. $M$ increases with $T$ at $T$$\,$$\approx$$\,$140$\div$150$\;$K for all the studied samples without signatures of splitting between ZFC and FC curves down to $T$$\,$$=$$\,$100$\;$K (exemplary results presented in Fig.$\;$\ref{FigFCZFCMagnVsTemp}).
Below 100$\;$K a bifurcation between ZFC and FC curves appeared, followed by a maximum in ZFC curve at around $T$$\,$$\approx$$\,$80$\div$90$\;$K. The obtained static magnetization results are in agreement with presented above ac susceptibility studies indicating clearly that we observe two magnetic transitions, e.g., paramagnet - cluster-ferromagnet at 140$\div$150$\;$K and reentrant cluster-ferromagnet - cluster-glass at 80$\div$90$\;$K in case of all the studied samples.

\subsection{High field results}

\begin{figure}[t]
  \begin{center}
    \includegraphics[width = 0.5\textwidth, bb = 0 40 720 640]
    {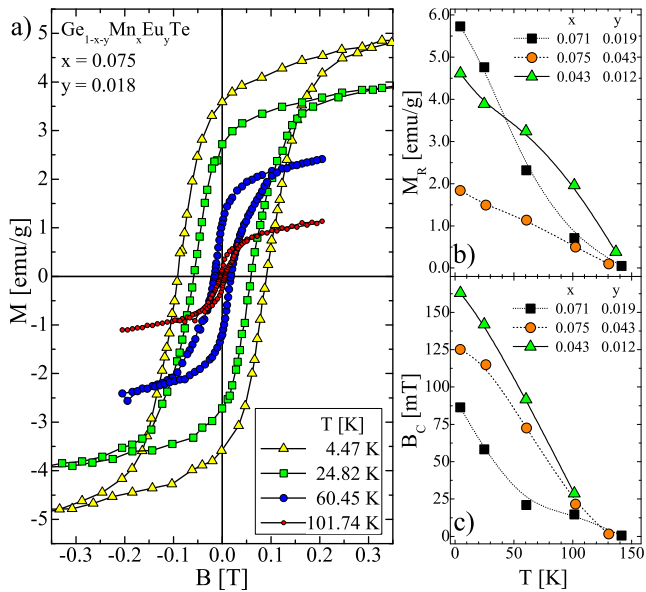}\\
  \end{center}
  \caption{\small (a) Magnetic hysteresis curves obtained for an exemplary Ge$_{0.907}$Mn$_{0.075}$Eu$_{0.018}$Te crystal and the temperature dependence of (b) remnant magnetization $M_{R}$ and (c) coercive field $B_{C}$ for selected Ge$_{1\textrm{-}x\textrm{-}y}$Mn$_{x}$Eu$_{y}$Te samples with different chemical composition.}
  \label{FigMHistCurves}
\end{figure}

\begin{figure}[t]
  \begin{center}
    \includegraphics[width = 0.42\textwidth, bb = 0 40 600 570]
    {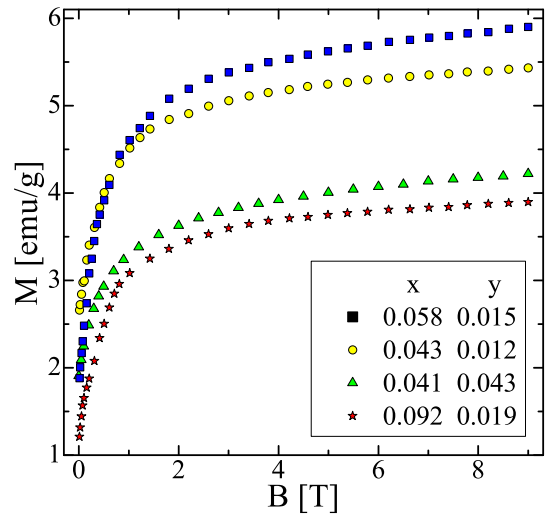}\\
  \end{center}
  \caption{\small Magnetization as a function of the applied magnetic field obtained at $T$$\,$$=$$\,$4.5$\;$K for selected Ge$_{1\textrm{-}x\textrm{-}y}$Mn$_{x}$Eu$_{y}$Te samples with different chemical composition.}
  \label{FigMHCurves}
\end{figure}

The magnetic field dependence of the magnetization shows a presence of well defined hysteresis curves at $T$$\,$$<$$\,$140$\;$K in the case of all our crystals (see Fig.$\;$\ref{FigMHistCurves}).
The hysteresis loops are observed at temperatures lower than 80$\;$K and therefore below the transition to the cluster-glass phase. That confirms our earlier conclusions, that in the cluster-glass phase the magnetic interactions of magnetic moments with a positive sign of the exchange constant are  dominant in our samples. The coercive field $B_{C}$ and remnant magnetization $M_{R}$ are very different in different samples (see Fig.$\;$\ref{FigMHistCurves}). However, we are not able to find any evident trends between both $B_{C}$ and $M_{R}$ and the chemical composition of the alloy. It is a signature that both $B_{C}$ and $M_{R}$ are rather closely related to the level of clustering of paramagnetic ions in the material than to the amount of Mn or/and Eu ions. \\ \indent The magnetic field dependence of isothermal magnetization is measured at higher magnetic fields, $B$$\,$$\leq$$\,$9$\;$T, and at temperatures lower than 200$\;$K with the use of Weiss extraction method employed to LakeShore 7229 magnetometer system. The obtained magnetization, $M$($B$), curves have similar shape in the case of all our samples. The exemplary $M$($B$) curves obtained at $T$$\,$$=$$\,$4.5$\;$K for selected Ge$_{1\textrm{-}x\textrm{-}y}$Mn$_{x}$Eu$_{y}$Te samples with different chemical composition are gathered in Fig.$\;$\ref{FigMHCurves}.
At temperatures above 150$\;$K the $M$($B$) curves have shapes typical of a paramagnetic materials. The shape of the $M$($B$) curves is nearly linear in magnetic field range 1$\,$$\leq$$\,$$B$$\,$$\leq$$\,$9$\;$T. It is a clear signature that aggregation of magnetic ions into clusters appears in the system.\cite{Anderson1990a} It is not possible, even with the use of relatively strong magnetic field, to saturate the magnetization of the samples. It shows that strong magnetic freezing occurs in the studied system.  It is not possible to obtain good fits of the experimental $M$($B$) curves to the Brillouin function. However, after addition of a phenomenological component linear with magnetic field to the $M$($B$) relation, reflecting the presence of magnetic clusters in a material, we obtain much better agreement between the experimental points and the fitted curves. \\ \indent The maximum values of the experimentally observed magnetization $M_{S}$ for our Ge$_{1\textrm{-}x\textrm{-}y}$Mn$_{x}$Eu$_{y}$Te samples are significantly lower than the values predicted by the Weiss theory by a factor 2$\,$$\div$$\,$4, depending on the alloy composition. It must be emphasized, that we do not know the real saturation magnetization value.

\section{Magnetotransport studies}\label{MagnetotransportStudies}

Basic characterization of the electrical properties of the Ge$_{1\textrm{-}x\textrm{-}y}$Mn$_{x}$Eu$_{y}$Te crystals consisted of resistivity and Hall effect measurements. The standard six contact DC Hall method is employed for the studies of electron transport. The measurements are done at room temperature with the use of magnetic field up to $B$$\,$$=$$\,$1.4$\;$T. Our results show, that the studied alloy has got low resistivity, $\rho_{xx}$$\,$$\approx$$\,$2$\times$10$^{-3}$$\;$$\Omega$$\cdot$cm, and a $p$-type conductivity with high carrier concentration, $n$$\,$$>$$\,$10$^{20}$$\;$cm$^{-3}$. The observed values of both carrier concentration and resistivity are typical for narrow gap GeTe, in which high carrier densities originate from large quantity of cation vacancies.\cite{Ure1960a}

\subsection{Magnetoresistance effects}

\begin{figure*}[t]
  \begin{center}
    \includegraphics[width = 0.65\textwidth, bb = 50 100 550 620]
    {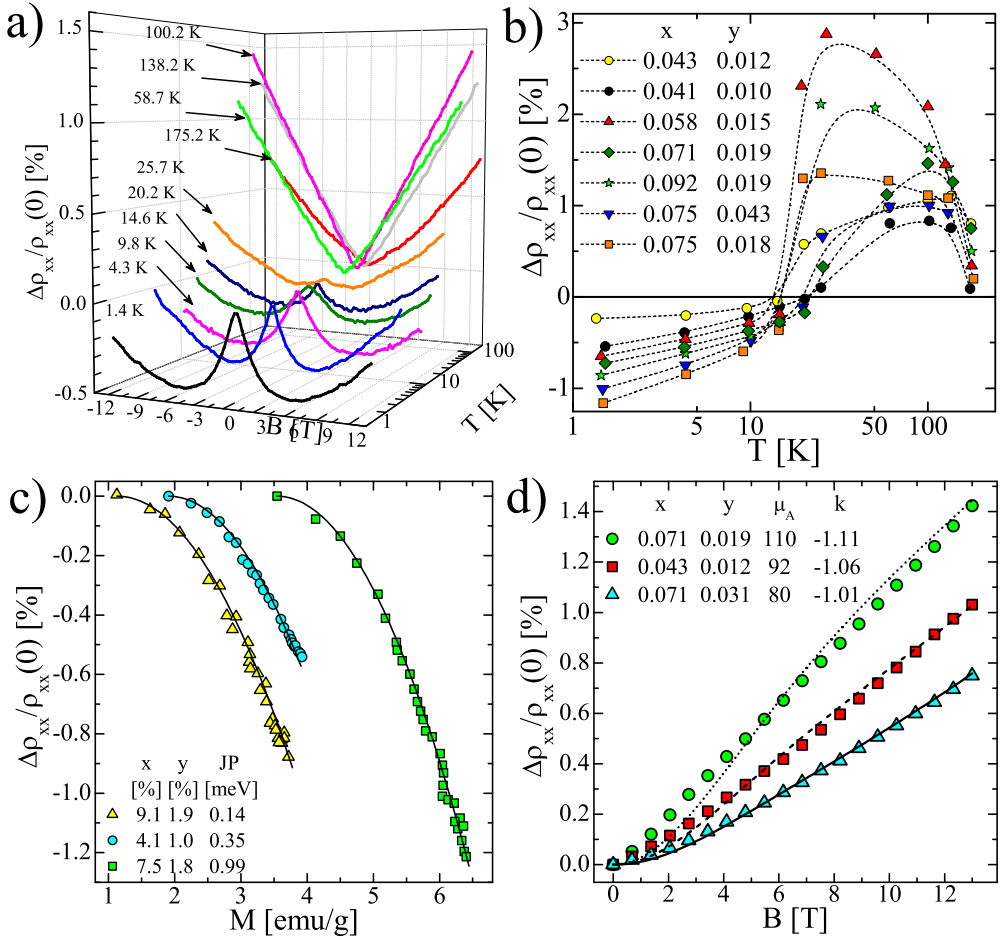}\\
  \end{center}
  \caption{\small Results of magnetoresistance measurements including: (a) magnetoresistance curves obtained at different temperatures for an exemplary Ge$_{0.910}$Mn$_{0.071}$Eu$_{0.019}$Te sample, (b) amplitude of the magnetoresistance as a function of temperature (points, the lines are added only to guide an eye), (c) the normalized resistivity $\Delta \rho_{xx}$/$\rho_{0}$ as a function of magnetization $M$ obtained experimentally at $T$$\,$$=$$\,$4.3$\;$K (points) and fitted theoretically (lines) and (d) the normalized resistivity $\Delta \rho_{xx}/\rho_{0}$ as a function of the magnetic field (experiment - points, theory - lines) at $T$$\,$$=$$\,$100$\;$K obtained in the case of Ge$_{1\textrm{-}x\textrm{-}y}$Mn$_{x}$Eu$_{y}$Te crystals with different chemical composition (see legend).}
  \label{FigMResCurves}
\end{figure*}

The DC measurements of the magnetoresistance (MR) are performed in the presence of magnetic fields $B$ up to 13 T. The obtained MR curves presented in the form of $\Delta \rho_{xx}/ \rho_{xx}(0)$$\,$$=$$\,$$(\rho_{xx}(B)$$\,$$-$$\,$$\rho_{xx}(B=0))$/$\rho_{xx}(B=0)$ curves show negative values at temperatures lower than 25$\;$K and positive values nearly linear with the magnetic field at 30$\,$$\leq$$\,$$T$$\,$$\leq$$\,$200$\;$K. The typically observed MR curves for the selected
Ge$_{1\textrm{-}x\textrm{-}y}$Mn$_{x}$Eu$_{y}$Te sample are presented in Fig.$\;$\ref{FigMResCurves}a.
The amplitude of the negative contribution to the MR (see Fig.$\;$\ref{FigMResCurves}b) increases with the amount of paramagnetic ions present in the sample from 0.2\% ($x$$\,$$=$$\,$0.043, $y$$\,$$=$$\,$0.012) up to 1.2\% ($x$$\,$$=$$\,$0.092, $y$$\,$$=$$\,$0.019). On the contrary, above $T$$\,$$=$$\,$30$\;$K only a positive and nearly linear MR with similar amplitudes is observed. It should be noted, that on the contrary to anisotropic MR, usually observed in homogeneous systems, all the MR effects observed in our material are isotropic. \\ \indent In the case of metallic ferromagnetic solids the negative isotropic MR is regarded as magnetic field induced decrease of carrier scattering in ferromagnetic clusters. Negative giant magnetoresistance has been reported in many inhomogeneous systems like Ni-SiO$_{2}$ alloys\cite{Gerber1997a} and is usually treated in framework of theory of spin-polarized electrons. The negative MR has been reported in numerous granular systems and is found to possess large values around 60\% (Ref.$\;$\onlinecite{Kilanski2010a}) as well as small values around 1\% (Ref.$\;$\onlinecite{Helman1976a}). The resistivity $\rho_{xx}$($B$,$T$) of a metal with ferromagnetic clusters can be expressed within the molecular field theory\cite{Helman1976a} using the approximate spin correlation function $m$ in the following form:
\begin{equation}\label{EqMr01}
   \frac{ \langle \overrightarrow{S_{1}} \bullet \overrightarrow{S_{2}} \rangle}{S^{2}} = m^{2}(B,T).
\end{equation}
The relative magnetoresistance $\Delta \rho_{xx}$/$\rho_{0}$ can be expressed with the use of the following equation:
\begin{equation}\label{EqMr02}
    \Delta \rho_{xx}/\rho_{0} = -\frac{JP}{4k_{B}T} \big{[}m^{2}(B,T) - m^{2}(B = 0, T)\big{]},
\end{equation}
where $P$ is the polarization of the tunneling electrons, $k_{B}$ is the Boltzmann constant, and $J$ is the electronic exchange coupling constant within the ferromagnetic grains. The above model reproduces the normalized magnetoresistance with only the intra-granular exchange constant $J$ and the carrier polarization $P$ as the fitting parameters, with the magnetization of the sample measured at the same temperature as the magnetoresistance. \\ \indent Both $J$ and $P$ constants in equation \ref{EqMr02} are not known for magnetically doped GeTe and cannot be separated. Therefore the fitting procedure gives the $J P$ product for each studied sample. The results presented in Fig.$\;$\ref{FigMResCurves}c show a good agreement between the experimental and fitted curves. The $J P$ product extracted from fits changes between 0.14 and 0.99$\;$meV. We find no correlation between the chemical composition of the samples and the values of $J P$. However, we do observe that $J P$ is proportional to the remanent magnetization $M_{R}$ of the samples. It is an important justification of the model, since the magnetization of the sample is strongly influencing the carrier polarization. The value of the carrier polarization constant is rather small not exceeding 1\% in granular metals such as Ni\cite{Campagna1976a} but can be much higher in MnAs with $P$$\,$$\approx$$\,$45\%.\cite{Panguluri2003a} Assuming $P$$\,$$\approx$$\,$1\% we obtain $J$ values higher than 100$\;$meV which is two orders of magnitude higher than the observed $k_{B} T$ values (around 2$\;$meV), at which the negative magnetoresistance disappears. It is then evident that the value of $P$ in our samples is at least as high as 15\%. Rather large changes of the $J P$ product for our samples might originate from the competing contributions of the $p$ and $d$ electrons to the magnetoresistance. \\ \indent The presence of a positive linear magnetoresistance is commonly attributed to the presence of inhomogeneities in a host material, having conductivity different than the host material. The linear, cluster-mediated magnetoresistance can have either large values of the order of 1000\% in MnAs-GaAs\cite{Johnson2010a} or values smaller than 1\% in case of thin Ni layers.\cite{Gerber1997a} The linear magnetoresistance has been also observed for IV-VI semimagnetic semiconductors.\cite{Brodowska2008b} An effective medium approximation (EMA) model describes quantitatively classical geometrical magnetoresistance of inhomogeneous media.\cite{Guttal2005a, Guttal2006a} The model assumes a fixed balance between phases A and B, called $p_{A}$ and $p_{B}$$\,$$=$$\,$1$-$$p_{A}$, respectively, and defined zero field conductivities $\sigma_{0A}$ and $\sigma_{0B}$. The effective conductivity of the material, $\sigma_{eff}$, can be calculated using the coupled self consistent equations, which can be expressed in a matrix form. The magnetic field dependence of resistivity can be calculated by solving the self consistent equation
\begin{equation}\label{EqLinMR01}
    \sum_{i=A,B} = p_{i} \delta \sigma_{i}(I-\Gamma\sigma_{i})^{-1} = 0,
\end{equation}
where $\Gamma$ is the depolarization tensor describing the geometrical properties of clusters and $I$ is the unity matrix. The EMA model\cite{Guttal2005a, Guttal2006a} is used to reproduce the experimental MR curves with given carrier mobility of the host matrix $\sigma_{A}$ (taken from low temperature Hall measurements) and the ratio between the carrier Hall constants $R_{HA}$ and $R_{HB}$ in the two phases A and B given by the parameter $k$ equal to $k$$\,$$=$$\,$$R_{HA}$/$R_{HB}$ being the fitting parameter. The MR curves calculated using EMA model are presented together with experimental data obtained at $T$$\,$$=$$\,$4.5$\;$K for the selected Ge$_{1\textrm{-}x\textrm{-}y}$Mn$_{x}$Eu$_{y}$Te samples having different chemical composition in Fig.$\;$\ref{FigMResCurves}d. We obtain good agreement between the theoretical and experimental curves in case of all the samples under investigation. The best fits to the experimental data are obtained for $k$ values around $k$$\,$$\approx$$\,$-1.1$\div$$\,$-1.0 indicating that the clusters present in the semiconductor matrix have the opposite ($n$-type) conductivity type.

\subsection{Anomalous Hall effect}

\begin{figure}[t]
  \begin{center}
    \includegraphics[width = 0.5\textwidth, bb = 103 130 740 510]
    {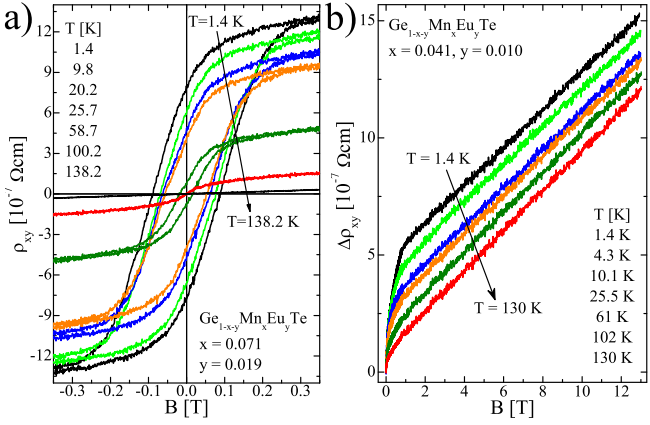}\\
  \end{center}
  \caption{\small The magnetic field dependence of the Hall resistivity $\rho_{xy}$($B$) obtained at different temperatures for the selected Ge$_{1\textrm{-}x\textrm{-}y}$Mn$_{x}$Eu$_{y}$Te samples with different chemical composition.}
  \label{FigAHE}
\end{figure}

\begin{figure}[t]
  \begin{center}
    \includegraphics[width = 0.42\textwidth, bb = 0 37 650 580]
    {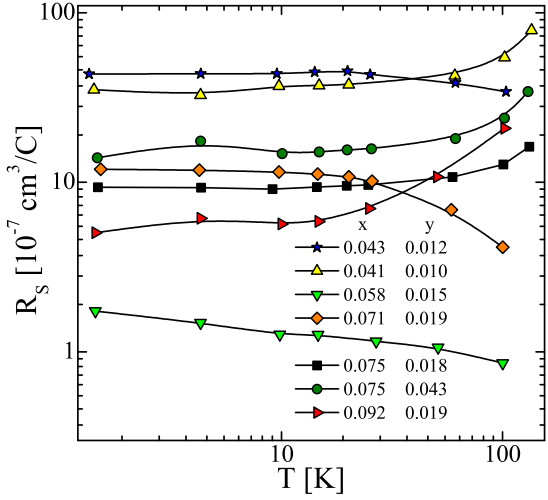}\\
  \end{center}
  \caption{\small The temperature dependence of the anomalous Hall effect constant $R_{S}$ for the Ge$_{1\textrm{-}x\textrm{-}y}$Mn$_{x}$Eu$_{y}$Te samples with different chemical compositions (see legend).}
  \label{FigRSvsT}
\end{figure}

The magnetic field dependence of the Hall effect is measured for all the investigated Ge$_{1\textrm{-}x\textrm{-}y}$Mn$_{x}$Eu$_{y}$Te samples in parallel with magnetoresistance measurements. The obtained isothermal Hall resistivity vs. magnetic field curves, $\rho_{xy}$($B$), at temperatures lower than 130$\;$K shows a significant anomalous contribution to the Hall effect (see Fig.$\;$\ref{FigAHE}).
Our results indicate the occurrence of well defined hysteresis loops in $\rho_{xy}$($B$) dependence (see Fig.$\;$\ref{FigAHE}), with the coercive field $B_{C}$ similar (within the limits of measurement accuracy) to that observed during magnetometric measurements. This result seems to be understandable because the anomalous Hall effect(AHE) depends on the magnetization of the material. \\ \indent Because of the strong AHE observed in the investigated material, the analysis of the data requires taking into account the effect of paramagnetic impurity scattering on the $\rho_{xy}$($B$) dependence. The data can be analyzed by separating the Hall effect into a normal $R_{H}$ and anomalous $R_{S}$ Hall constants, respectively, with the use of the following equation:
\begin{equation}\label{EqAhe01}
    \rho_{xy}(B) = R_{H} B + \mu_{0} R_{S} M,
\end{equation}
where $\mu_{0}$ is the magnetic permeability of vacuum. The second term in Eq.$\;$\ref{EqAhe01} describes the anomalous Hall effect which arises due to spin dependent scattering mechanisms. The anomalous Hall constant $R_{S}$ is determined from the experimental off-diagonal resistivity and magnetization data by the least square fit to Eq.$\;$\ref{EqAhe01}. The results of our calculations are presented in Fig.$\;$\ref{FigRSvsT}.
It may be seen that $R_{S}$ is almost temperature independent within the accuracy of its estimation in all our samples. It must be noted, that the observed changes in $R_{S}$($T$) dependence, observed in some of the samples at temperatures higher than 100$\;$K may be due to the difficulties in the data analysis at high temperatures, where AHE becomes small and it is more difficult to extract it from an ordinary Hall effect. It may be noted that with alloying the $R_{S}$ coefficient in the studied Ge$_{1\textrm{-}x\textrm{-}y}$Mn$_{x}$Eu$_{y}$Te crystals changed by an order of magnitude. The values of $R_{S}$ obtained in this work are similar to those reported for other IV-VI based diluted magnetic semiconductors.\cite{Dobrowolski2006a, Brodowska2006a, Kilanski2013a} The values of $R_{S}$ similar to those in earlier works suggest, that AHE in the studied material does not seem to be correlated with the presence of magnetic clusters in Ge$_{1\textrm{-}x\textrm{-}y}$Mn$_{x}$Eu$_{y}$Te samples. However, one can distinguish within the data a general trend towards a decrease of $R_{S}$ with an increasing amount of Mn and Eu. It seems that the value of $R_{S}$ in the studied samples is much more complex function of parameters other than chemical composition. The carrier concentration, $n$, and the coercive field, $B_{C}$, play the important role in carrier scattering in ferromagnetic materials. The presence of AHE is widely observed in many systems covering both homogeneous systems like Ga$_{1\textrm{-}x}$Mn$_{x}$As (Ref.$\;$\onlinecite{Jungwirth2006a}) and cluster ferromagnetic semiconductors like Zn$_{1\textrm{-}x}$Co$_{x}$O (Ref.$\;$\onlinecite{Hsu2008a}) and was analyzed in the way presented above. \\ \indent A more detailed analysis of the magnetotransport data can be done in order to determine the major scattering processes leading to the appearance of AHE in a material. There are two major scattering mechanisms involved in AHE in metallic ferromagnets, i.e., the skew scattering\cite{Smit1955a} and side jump\cite{Berger1970a} effects. The scaling analysis of the experimental data is performed with the use of an exponential scaling relation  modifying equation \ref{EqAhe01} into the following relation:
\begin{equation}\label{EqAhe02}
    \rho_{xy}(B) = R_{H}B + c_{H} \rho_{xx}^{\delta} M,
\end{equation}
where $c_{H}$ is the scaling constant and $\delta$ is the scaling exponent. It is well known, that AHE resulting purely from either skew scattering or side jump processes should give a scaling factor $\delta_{H}$ equal to 1 or 2, respectively. The performed least square fits of equation \ref{EqAhe02} to the magnetotransport and magnetization data allow us to estimate the temperature dependencies of $\delta$ for all the studied crystals. The results show that all the values of $\delta$ fall in the range from 1.1 and 1.3$\pm$0.1. This implies that the skew scattering mechanism is the most important scattering mechanism responsible for the observed AHE in this system. However, since $\delta$ is not equal to 1, it is highly probable that effects arising from side jump mechanism play also a role in the AHE in Ge$_{1\textrm{-}x\textrm{-}y}$Mn$_{x}$Eu$_{y}$Te crystals.

\section{Summary}\label{Summary}

To conclude, we report the experimental studies of magnetic and magnetotransport studies of Ge$_{1\textrm{-}x\textrm{-}y}$Mn$_{x}$Eu$_{y}$Te crystals. Magnetic order is observed at $T$$\,$$<$$\,$150$\;$K for all the studied samples and can be interpreted as a cluster ferromagnetism at 90$\,$$<$$\,$$T$$\,$$<$$\,$150$\;$K and the cluster-glass ordering at $T$$\,$$<$$\,$90$\;$K. The magnetic properties of the alloy show that the transition temperatures do not depend on the amount of paramagnetic ions in the alloy. \\ \indent Ferromagnetic clusters have a substantial impact on magnetotransport properties of the alloy. The presence of negative magnetoresistance at $T$$\,$$\leq$$\,$25$\;$K caused by quantum tunneling of carriers is observed in all the studied crystals. Moreover, at temperatures 25$\,$$<$$\,$$T$$\,$$<$$\,$200$\;$K a positive magnetoresistance linear with $B$ is observed, a geometrical effect characteristic for inhomogeneous systems. Both effects are reproduced in a satisfactory manner by means of appropriate theoretical models of magnetoresistance in inhomogeneous materials. Thanks to europium alloying in Ge$_{1\textrm{-}x}$Mn$_{x}$Te it is possible to observe gathering of paramagnetic ions into ferromagnetic clusters, and observe new, interesting magnetotransport effects in this material. \\ \indent A strong, temperature independent AHE is observed in all the studied crystals at temperatures lower than the Curie point. Large changes of the $R_{S}$ factor between the samples are attributed rather to a difference in a domain structure and carrier concentration in these crystals than to a difference in the amount of paramagnetic ions. The scaling analysis of the experimental data shows, that AHE in the studied alloy is mainly due to the skew scattering processes.

\section{Acknowledgements}\label{Acknowledgements}

\noindent The research was partly supported by the Foundation for Polish Science - HOMING PLUS Programme co-financed by the European Union within European Regional Development Fund. The research was partly financed by the National Center for Science of Poland under decision number DEC-2012/05/D/ST3/03161


\end{document}